# BUSINESS INTELLIGENCE AND SUPPLY CHAIN AGILITY

**Mohammad Moniruzzaman**
Department of Computing and Information Systems
The University of Melbourne
Parkville, Victoria, Australia
Email: monir.australia@gmail.com

**Sherah Kurnia**
Department of Computing and Information Systems
The University of Melbourne
Parkville, Victoria, Australia
Email: sherahk@unimelb.edu.au

**Alison Parkes**
Department of Computing and Information Systems
The University of Melbourne
Parkville, Victoria, Australia
Email: aparkes@unimelb.edu.au

**Sean B. Maynard**
Department of Computing and Information Systems
The University of Melbourne
Parkville, Victoria, Australia
Email: sean.maynard@unimelb.edu.au

## Abstract

Supply Chain Agility is vital for organisations wanting to remain competitive in today's dynamic business environment. There is increasing interest in deploying Business Intelligence (BI) in the Supply Chain Management (SCM) context to improve Supply Chain (SC) Agility. However, there is limited research exploring BI contributions to SC Agility. In this research-in-progress paper we propose a model based on a conceptual analysis of the literature showing how BI can help organisations achieve SC Agility by supporting the key areas of SCM (Plan, Source, Make, Deliver, and Return). In the next stage of this project, we will conduct a series of case studies investigating how organisations use BI when managing their SC activities and how BI contributes to SC Agility. The result of the study will help organizations deploy BI effectively to support SCM and improve SC Agility.

**Keywords:** Business Intelligence, Agility, Resource-based View, Supply Chain, Supply Chain Management

## 1   INTRODUCTION

A single organization is unlikely to have the full capability of bringing products or services to the market to meet changing customer demands. Organizations are frequently forced to form a network to work collectively to meet customer demands. Christopher (1999) refers to such a collective network as a Supply Chain (SC), which is responsible for the planning and management of all activities involved in acquisition, conversion, and delivery of products/services to consumers. SC participants need to work closely together to meet consumer demands rather than competing against each other. Therefore, the nature of competition has shifted from 'organization against organization' to 'Supply Chain against Supply Chain' (Ketchen et al. 2007). In today's dynamic environment, flexibility is critical in managing SC so that organisations can deal with market changes and convert these changes into opportunities (Agarwal et al. 2007). The concept of Agile SC has been introduced to define the SC capability that enables an organization to respond to unpredictable changes and uncertainties in dynamic business environment. Achieving SC Agility is challenging and has become a research topic of increasing interest over the last decade.

Whilst some research exists (Swafford et al. 2008; Wu et al. 2006; Yusuf et al. 2014a) that investigates the contributions of IT tools in achieving Agility in SC, how Agility in SC can actually be achieved by organisations through the use of IT is still unclear. As most studies explore IT holistically, not enough evidence of the utility of specific IT components is provided. In particular, studies assessing the contribution of Business Intelligence (BI) in helping organisations achieve SC Agility are lacking. Furthermore, the majority of studies (Christopher 2000; Yusuf et al. 2004) focus on either 'Agility enablers' or partial aspects of Agility in SC to consider IT contributions, whereas Agility is a complex concept with multiple dimensions. This level of complexity requires a thorough study of all aspects of Agility to enhance the current understanding. In most research SCM is being considered only at a high level instead of addressing the key areas of planning, sourcing, making, delivering and returning goods. Achieving Agility requires integrated management of all aspects of SC activities, a complete understanding of BI contributions can only be obtained by considering all aspects of SCM in one study. Furthermore, while some studies investigate the role of BI in the context of SCM performance (Oliveira et al. 2012; Trkman et al. 2010a), none has assessed BI contributions to Supply Chain Agility. Finally, most existing studies report survey data which limits the richness and understanding of how BI can contribute to SC agility. Additional rich case studies would improve current understanding of this complex and important area.

To address current knowledge gaps this study explores the contributions of BI use in SCM towards achieving Agility in SC. It examines the use of BI in SCM key areas and explores whether this BI use can help improve Agility in SC. The overall research question for this study is:

*How does BI-enabled Supply Chain Management improve Supply Chain Agility?*

In this research-in-progress paper, we review the literature in the areas of Supply Chain Management and Business Intelligence to identify how BI can potentially contribute to SC Agility. We adopt a Resource Based View and use Organizational Information Processing Theory to conceptualize possible mechanisms to achieve SC Agility; we build in these foundations and present a conceptual research model and propositions.

The outline of this paper is as follows. In the next section, we explain our key concepts and establish the theoretical basis of the work. Next we present our conceptual research model and propositions. To conclude we outline the next steps of the study and identify the expected contributions of this research.

## 2   BACKGROUND

This section reviews three key areas of literature: supply chain management, supply chain agility, and business intelligence. It then presents literature that has discussed SC Agility and SCM, before presenting the identified knowledge gaps in the literature.

### 2.1   Supply Chain Management

New et al. (1995) defines Supply Chain as a collection of organizations participating in a network to ultimately meet market demands and illustrates the SC cycle as depicted in Figure 1. A typical Supply Chain activity starts from raw material extractions from nature to produce raw materials for component manufacturer. Final product manufacturer produces final product using components and wholesale to retailer to ultimately deliver to final consumers. Distribution and warehousing functions support the whole chain in appropriate stages.

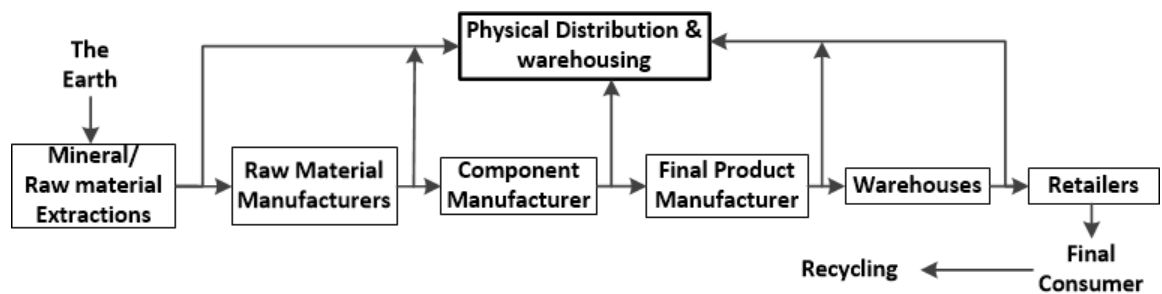

*Figure 1: Supply Chain processes (New et al. 1995)*

There are various business processes within a typical Supply Chain that share the common objective of satisfying changing customer demand. Based on the Supply Chain Operations Reference (SCOR) model, the processes involved in SCM can be categorized into five groups; Plan, Source, Make, Deliver and Return(Apics 2015). Business processes across these five areas need to be coordinated and managed to ensure products and information flow efficiently within Supply Chains. Therefore, the concept of Supply Chain Management (SCM) has been introduced since 1990s (Lambert et al. 1998).

SCM involves a collection of integrated processes responsible for the acquisition of raw material, production and distribution of final products/services to satisfy end user need (Houlihan 1985; Mentzer et al. 2001). Within the context of SCM, IT contributions have been well acknowledged to help organisations improve information flow and coordinate various activities efficiently and effectively {Simchi-Levi, 2008 #425}{Simchi-Levi, 1999 #424}.

## 2.2 Agility in SC

SC agility is the ability of an organization to reconfigure, adjust, and change resources for the key SC processes (Plan, source, make, deliver and return) to satisfy the demand of a changing environment (Agarwal et al. 2007; Lin et al. 2006a; Yusuf et al. 2014a). Extant research (Braunscheidel et al. 2009; Christopher 2000; Giachetti et al. 2003; Lin et al. 2006a; Xun et al. 2009) identifies four key elements of SC Agility: Flexibility, Responsiveness, Competency and Quickness, these are now described in detail.

Flexibility is the ability of a SC to cope with change. Christopher (2000) identified flexibility as an important feature of agility that is needed to fulfil highly uncertain and varied market demand. Flexibility describes the SC's ability to cope with both certain and uncertain changes by reconfiguring resources and processes to meet changes in customer demands (Braunscheidel et al. 2009; Gong et al. 2012; Xun et al. 2009).

Responsiveness refers to the SC's ability to respond to a dynamic market. Unexpected demand requires a quick assessment of organization capabilities to determine if demand can be meet and provide feedback to consumers within short timeframes (Xun et al. 2009). An integrated view of the overall supply status is critical to be able to respond to customers in a timely manner (Braunscheidel et al. 2009; Qrunfleh et al. 2013).

Competency is the SC's ability to achieve its goals. Ngai et al. (2011) identify 3 components of SC competence which include: "IT Integration and flexibility", "operational competence" and "management competence". IT Integration facilitates information flows within SC to forecast change and address change (Zhang et al. 2013). Operational Competence refers to the ability of a SC to utilize its resources to deliver according to market demand(Braunscheidel et al. 2009). Finally, Management Competence refers to the role and vision of top management and the level of employee skills.

Quickness is the ability of a SC to execute activities within the shortest possible timeframe. The SC needs to continuously adjust and restructure its relevant processes and strategies to respond quickly to market changes (Agarwal et al. 2007). Increasing sophistication in customer preferences means highly customized products are often required (Chen et al. 2011a; Power et al. 2001) so a SC needs to be able to deliver customized products within short lead times without compromising quality or cost (Lin et al. 2006a; Ngai et al. 2011; Stratton et al. 2003).

In summary, an agile Supply Chain has four characteristics: responsiveness, flexibility, competence and quickness ( Braunscheidel et al. 2009; Christopher 2000; Huang et al. 2009; Ngai et al. 2011; Sherehiy et al. 2007; Xun et al. 2009; Yusuf et al. 2004). Through this study, we explore how BI can support SCM practices and achieve these four characteristics of Supply Chain Agility.

## 2.3 The potential of Business Intelligence (BI)

Isik et al. (2011) define Business Intelligence (BI) as a mix of business and technical elements that pull 'historical data' from internal/external sources and transform it into meaningful information for management decision making. Business use of BI refers to a tool that analyse data and support decision making (Trkman et al. 2010b) while the technology aspects deal with how required BI infrastructure can be made available. This study considers BI as a query and reporting tool that is relevant for decision making in dynamic market environment.

Researchers identify three major uses of BI. First, BI integrates a wide range of data into a single repository, providing a single enterprise-wide view via management dashboards and reports. (Long et al. 2012). Second, BI transforms data into information that is valuable for managers to evaluate options (Acar et al. 2010). Third, BI delivers the capacity to identify the root cause of problems by providing information drill down (Hocevar et al. 2010).

BI provides insights, based on data analysis, to assist managers in decision making within highly changing environments (Isik et al. 2013). Shanks et al. (2010b) established that BI use in an organization assists managers in taking value creating actions by relying on facts or information provided by BI query and reporting. Although there have been some studies assessing the impact of BI use in SC performance (Trkman et al. 2010a), there has been limited systematic and detailed research showing how BI assists organizations in Supply Chain Management.

## 2.4 Existing studies related to BI, SCM and SC Agility

This section synthesizes relevant literature and analyses gaps in the existing literature. To find existing literature around Agile SC, we used Discovery and Google Scholar, searching using different combinations of keywords: "Agile Supply Chain", "Information Technology", "Business Analytics" and "Business Intelligence". Our search identified 357 papers mentioning Agile SCM. We also explored related literature using forward citations in the key papers. We read through the titles and abstracts of all papers to identify those papers that discuss IT/BI contributions in Agile SC, resulting in twenty papers (Table 1).

Table 1 summarizes five specific aspects addressed in the shortlisted prior studies in the area of Agile SC. None of the 20 papers summarised in table 1 address all of key concepts of Business Intelligence, Supply Chain Management, and Supply Chain Agility. Five of the papers address at least one of the four SC agility characteristics, while another two consider at least one of the supply chain management key process areas. Four papers consider the relationship between one or more of the SCM key process areas and SC agility characteristics, another four papers consider relationships between one or more of the SC management key process areas and Business Intelligence. The remaining five papers are included as they contain general perspectives on the relationship between SC agility and SC management. However, none of these papers specifically considers Business Intelligence.

| Relevant Literature | Business Intelligence focus | SCM areas | SC Agility | Research Method | Location |
|---|---|---|---|---|---|
| (Ngai et al. 2011) | Not BI specific | SC in general | Flexibility, Competence | Multiple case studies | Hong Kong |
| | **Findings**: Considered IT as an important component of SC Agility. SC Agility positively impacts organization's performance. | | | | |
| (Swafford et al. 2008) | Not BI specific | SC in general | Flexibility, Quickness | Survey | USA |
| | **Findings**: Positive role of IT integration in SC Flexibility. SC Agility directly improves competitive performance. | | | | |
| (Liu et al. 2013) | Not BI specific | SC in general | Responsiveness, integration competency | Survey | China |
| | **Findings**: IT capability affects organization's performance through impacting "absorption capability and SC Agility". | | | | |
| (Cheng et al. 2011) | Partial | Make | Competence | Survey | Hong Kong |
| | **Findings**: IT integration positively impacts organization's performance. | | | | |
| (Chakraborty et al. 2011) | Not BI specific | SC in general | Flexibility, Responsiveness | Survey | India |
| | **Findings**: Identified 7 key factors of SC Agility: Technology, partnership, quality, education, market, competence, and team building. | | | | |
| (Dowlatshahi et al. 2006) | Not BI specific | Make | Agility enablers | Survey | USA |
| | **Findings**: As enabler of Agile manufacturing, both Virtual enterprise and IT positively impact organization performance. | | | | |
| (Hasan et al. 2012) | BI focus | Make | Agility enablers | Single Case Study | India |

| Relevant Literature | Business Intelligence focus | SCM areas | SC Agility | Research Method | Location |
|---|---|---|---|---|---|
| | **Findings**: Positive contribution of Analytical Network Processing in Agility in Production. | | | | |
| (Wu et al. 2006) | Not BI specific | SC In general | Agility in general | Survey | USA |
| | **Findings**: Positive Impact of IT in SC Agility and organization performance. | | | | |
| Yusuf et al. (2014b) | Not BI specific | SC in general | Agility enablers | Survey | UK |
| | **Findings**: Identified dimensions and attributes of Agile SC. | | | | |
| Yusuf et al. (2014a) | Not BI specific | SC in general | Agility enablers | survey | USA |
| | **Findings**: Positive role of cluster strategy in Agility | | | | |
| Qrunfleh et al. (2013) | Not BI specific | Sourcing | Responsiveness | Survey | USA |
| | **Findings**: SC responsiveness impact organization's performance | | | | |
| Bottani (2010) | Not BI specific | SC in general | Agility enablers | Survey | Europe |
| | **Findings**: 8 Agile attributes and 4 Agile enablers has been identified | | | | |
| Trkman et al. (2010a) | BI focus | Plan, Source, Make, Deliver | N/A | Survey | USA, Europe, Canada, Brazil, and China. |
| | **Findings**: Positive impacts of analytic type of BI in SCM with mediating impact from information systems and process orientation. | | | | |
| Oliveira et al. (2012) | BI focus | Plan, Source, Make, Deliver | N/A | Survey | World-wide |
| | **Findings**: Impact of analytic type of BI in SCM is influenced by the business process maturity of an organization. | | | | |
| Yusof et al. (2013) | BI focus | Make | N/A | Literature review | N/A |
| | **Findings**: Affirms positive contribution of BI in manufacturing processes. | | | | |
| (Braunscheidel et al. 2009) | Not BI specific | SC in general | Responsiveness Competency Flexibility | Survey | USA |
| | **Findings**: Beside flexibility, integration among participants in a supply chain plays vital role in achieving Agility in SC. | | | | |
| (Xun et al. 2009) | Not BI specific | SC in general | Dimensions of SC Agility | Survey | USA |
| | **Findings**: Identified six dimensions to measure SC Agility: strategic alertness, strategic response capability, operational alertness, operational response capability, episodic alertness and episodic response capability. | | | | |
| (Chen et al. 2011b) | Not BI specific | Return, Deliver | N/A | Analytic model | N/A |
| | **Findings**: Return can be managed through supply chain configuration considering manufacturer and retailer agree on buy back prices for unsold inventory and customer return. | | | | |

| Relevant Literature | Business Intelligence focus | SCM areas | SC Agility | Research Method | Location |
|---|---|---|---|---|---|
| (Xiao et al. 2010) | Not BI specific | Return, Deliver | Responsiveness, flexibility | Analytic model | N/A |
| | **Findings**: Refund amount derived from return plays an important role in decision making and profitability of SC participant. | | | | |
| (de Brito et al. 2009) | Not BI specific | Return, sourcing | Flexibility, responsiveness | Analytic model | N/A |
| | **Findings**: Information quality plays critical role in return forecast using return information system and inventory control. | | | | |

*Table 1: Summary of selected studies on SCM and SC Agility*

## 2.5  Research Gaps

Based on the literature analysis and synthesis conducted, five research gaps have been identified in the existing studies around BI/IT, SCM and SC Agility.

*First*, an analysis of coverage of the four supply chain agility characteristics (flexibility, responsiveness, competence, quickness) showed that 11 papers did not address these specific characteristics, focusing instead on agility enablers (seven papers) or not considering agility (four papers). Of the nine papers which addressed the supply chain agility characteristics none considered all four. Two papers looked at only one characteristic, six papers considered two, and one paper considered three of these four dimensions. Coverage of the characteristics was also uneven with six papers considering flexibility and six including responsiveness. Four papers considered competence while only one paper looked at quickness. We argue that to understand contributions of SC Agility on competitive advantage in the market, a complete study focusing on all key characteristics of Agile SC need to be considered (Yusuf et al. 2014a).

*Second*, considering the Supply Chain management key process areas (plan, source, make, deliver, return) 10 of the 20 papers did not address any of these key process areas, concentrating instead on more generic views of SCM. Of the 10 papers which did address these key process areas none considered all five areas. Five papers considered only one process area, three papers considered two process areas, and the remaining two papers considered four of the five process areas. In terms of coverage of each process area two papers considered plan, four included source, 6 considered make, four included deliver and three involved return. To achieve Agility, end to end SC needs to perform in an integrated way. Hence, the role of all key SCM process areas need be considered in SC Agility related research (Huang et al. 2009; Lin et al. 2006a).

*Third*, Business Intelligence coverage was similarly sparse with only four papers focusing directly on BI and one paper partially addressing it. The remaining 15 papers did not consider business intelligence specifically instead they looked at information technology generically. Again, 5 papers invested the role of BI in the context of Supply Chain Management, none of them has assessed BI contributions to the achievement of Agile Supply Chain. Trkman et al. (2010a), for example, studied SC operations in different organizations and found that Business Intelligence contributes to SC performance with moderating effect from Information systems and business processes. However, Trkman et al. (2010a) did not relate BI with SC Agility components. On the other hand, de Brito et al. (2009) focused on return forecasting and acknowledged the value of information in managing the Return process. However, their research didn't include BI as a potential tool to support return data analysis. Xiao et al. (2010) and Chen et al. (2011b) recognise that SC flexibility and responsiveness can be achieved through the coordination between the manufacturer and the retailer with respect to managing risk due to returns. However, neither study acknowledged that the analysis of the integrated view of return historical data analysis could be delivered by a BI tool (Davenport 2006; Sahay et al. 2008).

*Fourth*, a wide range of research methods were used in the paper reviewed. Fourteen papers used surveys, one paper was not empirical, and three used analytical modelling techniques. The remaining two papers contain case studies, one paper analysed multiple cases and the other contained a single case. Most of the studies employ a survey approach with limited studies using a qualitative approach.

Though a survey helps researchers to have data from a number of different organizations, the data may not provide a complete understanding about SC agility as the concept of Agility in SC is relatively new and perception based (Braunscheidel et al. 2009). On the other hand, case studies allow for in-depth understanding to be obtained regarding how BI contributes to the achievement of SC Agility through direct interactions with subject organization resources (Yin 2011). Moreover, a case study is the most appropriate method for 'how and why' type of research questions (Yin 2013).

*Fifth,* as shown in Table 1, most of the studies are based on data from the USA, China, or India. Little work has been done in Australian or New Zealand contexts. We argue that due to the geographic location of Australia/New Zealand, the distance to global suppliers, logistic costs and lead time are different to other parts of the world. Therefore, how BI can contribute in SC Agility in an Australian or New Zealand context may be different to other countries.

Based on the knowledge gaps identified, our study aims to specifically explore how BI creates SC Agility by considering the use of BI in the main areas of SCM. To gain an in-depth understanding, we will conduct a qualitative study involving multiple case study organisations in the Australian and New Zealand region.

## 3 Theoretical foundation

The Resource Based View (RBV) offers an appropriate lens to this study to understand how BI supported SCM can improve supply chain agility. The effectiveness of BI also depends on how BI is used within SCM processes. Organization information processing theory (OIPT) shows that information resources can be implemented in an organization design to deal with uncertainty. Thus RBV and OIPT will be used to build a solid framework for this study to explore how BI enabled SCM can influence Agility in SC. Each theory is explained in more detail in following subsections.

### 3.1 Resource Based View (RBV)

The RBV considers organizations' assets, processes, information, know-how, capabilities, etc. as resources. These resources contribute to competitive advantage if they possess two attributes: resource heterogeneity and resource immobility (Barney 1991). Barney (1991) explains heterogeneity attribute of organization's resource(s) using two terms: valuable and rare. Valuable attribute of a resource refers to the ability to outperform competitors or reduce organization's own weaknesses. Rare refers to exceptional or uncommon feature(s) of a resource so that competitor can't conceive the same. Valuable and rare resources of an organization can be a source of competitive advantage as long as competitors are unable to obtain such resources. Barney (1991) refers to this condition as 'Organization Resource Immobility' which has two aspects of resources: in-imitable and Non-substitutable.

A number of studies confirm that BI capability is a valuable resource for organizations, as BI can improve organizational performance (Davenport 2006; Sahay et al. 2008; Trkman et al. 2010a). BI capability embedded in business processes can become a rare capability of an organization, as organization specific processes are not easily available for competitors in the industry (Barney 2001; Isik et al. 2013). BI embedded business processes are hard for other organizations to duplicate as they are specific to the culture and management skill set of an organization. This complies with the in-imitable requirement of a resource as suggested by RBV. Thus, with the support from RBV theory, this study argues that BI can be a Valuable, Rare, In-imitable and Non-substitutable resource for an organization, particularly in SC functions. Davenport (2006) further argues that organizations using BI may experience a performance increase, but not necessarily achieve competitive advantage. Therefore, it is important to explore how BI can be effectively used in business processes to gain competitive advantage in the market.

### 3.2 Organization Information Processing Theory (OIPT)

OIPT was proposed by Galbraith (1973) in the context of organizational design strategy. OIPT explains the linkage between information as resource and the use of information resources as an important factor for organization performance (Fairbank et al. 2006). In the course of organization design, Galbraith (1973) views organization as a collection of tasks and subtasks that need to be integrated to the whole tasks to ensure organization performance. Task owners within organizational hierarchy need to deal with 'task uncertainty' due to the novelty of tasks involved or inadequate knowledge to perform the tasks (Tatikonda et al. 2000). To deal with task uncertainty Galbraith (1973) proposed two sets of non-mutually exclusive strategies: (1) reducing the need for information processing and (2) increasing the capacity to process information. An example of the first strategy is to invest in buffer stock to deal with demand/supply uncertainty in Supply Chain (Premkumar et al. 2005). An example of the second strategy is to build BI capability to make relevant information available to task performer so that appropriate decision can be made to deal with task uncertainty.

IS research provides empirical evidence in favour of the increased information processing strategy suggested by OIPT. Fairbank et al. (2006) conducted a longitudinal research study within the U.S life/health insurance industry. Based on 403 responses, they found that organizations which using information processing can perform better than compitotors in the market. Premkumar et al. (2005) conducted a study involving 169 managers from publicly traded organizations finding that 'Supply Chain Intelligence' and 'Marketing Intelligence' play a critical role in the organization's performance in new product development.

Information processing need is correlated with the level of task uncertainty. If planned targets are not met at a management function level then additional decisions may be required at higher levels of the management hierarchy. Hence, more cross functional information is required to be processed at the higher level of hierarchy. Thus OIPT proposes that organizations need to be "structured around information and information flows in an effort to reduce uncertainty" (Fairbank et al. 2006). Based on our understanding of the OIPT, we argue that a BI capability needs to exist in each task / subtask / hierarchy of SCM processes so that the required data is available to respective task owners to make decisions and optimize performance. Such implementation of BI capability may offer respective organization a unique bundle of resources that are in-imitable and un-substitutable.

## 4    RESEARCH MODEL AND PROPOSITION DEVELOPMENT

In this section we present a research model and propositions, based on RBV theory and the OIPT, showing how BI can be used within the five areas of SCM to improve SC Agility. SC Agility is a capability considered as a strategic resource to help an organization achieve competitive advantage (Christopher 1999; Ivanov 2010; Qrunfleh et al. 2013; Stratton et al. 2003; Yusuf et al. 2014a). Our conceptual research model is based on the theoretical foundations of RBV and OIPT as discussed previously. Figure 2 shows our proposed research model and Table 2 provides the definition of each construct in the model:

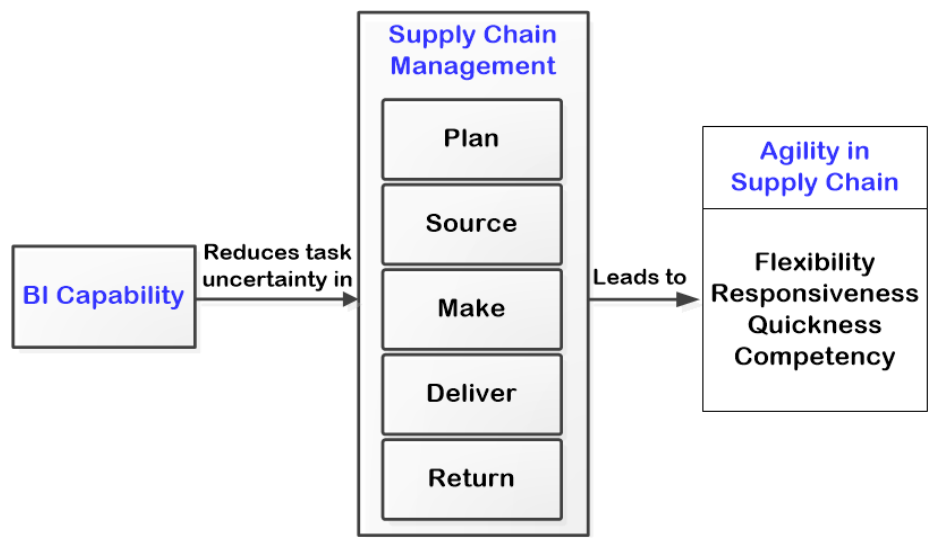

*Figure 2- Conceptual research model*

In the conceptual research model, BI is considered as a capability of an organization while SCM activities are considered as organization processes. Applying the RBV, this model posits that BI can be integrated in SCM key process areas such as Plan, Source, Make Deliver, and Return to form a resource bundle which can support Supply Chain managers in decision making. BI supported SCM potentially can help organizations in achieving another capability - SC Agility, noting that integration of BI may have many forms. With the support from OIPT, we argue that BI can be integrated in SCM processes according to information need and 'task uncertainty' involved at different levels of organization hierarchy.

| Definition | References |
| --- | --- |

| Definition | References |
|---|---|
| **BI Capability:** The ability of an organization to integrate multiple data sources and present summarized information providing insight to users. BI capability also supports what-if analysis and predictive analysis based on historical data patterns. | (Davenport 2006; Isik et al. 2011; Ramakrishnan et al. 2012; Ranjan 2008; Sahay et al. 2008; Sharma et al. 2010). |
| **Supply Chain Management:** A set of integrated processes that is responsible for acquisition of raw material, production, and distribution of final products/services to satisfy end user need. | (Christopher 1999; Mentzer et al. 2001; Soni et al. 2013; Zhou et al. 2011) |
| - **Plan:** A SCM function that deals with preparing actions for effective source, make, deliver and return processes. | (Danese 2011; Gupta et al. 2003; Ivanov et al. 2012) |
| - **SC Source:** A SCM function that acquires resources required for the 'make' processes. | (Christopher 1999; Tan 2001; Zhou et al. 2011) |
| - **SC Make:** A SCM function that transforms raw materials into product to satisfy market demand. | (Cagliano et al. 2004; Vinodh et al. 2013; Yusof et al. 2013; Yusuf et al. 1999) |
| - **SC Deliver:** A SCM function that hands over ordered products to customers. | (Christopher 1999; K et al. 2009; Yang 2013) |
| - **SC Return:** A SCM function that collects used or defective products from customers. | (Chen et al. 2011b; de Brito et al. 2009; Xiao et al. 2010; Yang 2013; Zhou et al. 2011) |
| **Agility in SC:** The capability of a Supply Chain to respond to dynamic market conditions with the following four characteristics. | (Christopher 2000; Christopher et al. 2001; Huang et al. 2009; Lin et al. 2006a; Liu et al. 2013; Power et al. 2001; Swafford et al. 2008; Yusuf et al. 2004 |
| - **Flexibility:** the capability to reconfigure resources and processes quickly to satisfy customer demand. | (Giachetti et al. 2003; Gong et al. 2012; Swafford et al. 2008; Wadhwa et al. 2008; Xun et al. 2009) |
| - **Responsiveness:** the capability to predict uncertainty and respond before competitors. | (Braunscheidel et al. 2009; Power et al. 2001; Yusuf et al. 2004; Yusuf et al. 1999) |
| - **Quickness:** the capability to perform an activity within the shortest possible timeframe. | (Chakraborty et al. 2011; Christopher 1999; Christopher et al. 2001; Liu et al. 2013) |
| - **Competency:** the capability to achieve SC goals. | (Cagliano et al. 2004; Huang et al. 2009; Lin et al. 2006a; Naim et al. 2011; Ngai et al. 2011; Swafford et al. 2006; Yusuf et al. 2014a; Yusuf et al. 2014b) |

*Table 2: Key concepts definitions for the current study*

BI has the potential to reduce task uncertainties by providing insights in the form of multidimensional sales forecast / analysis reports, multidimensional delivery analysis reports, and cost analysis reporting. This functionality supports make or buy decisions, demand forecast reporting to facilitate production scheduling, and so on (Davenport 2006; Ranjan 2008). Next we explain how BI can be used to support the key SCM process areas including Plan, Source, Make, Deliver, and Return.

The first SCM key area, *'plan'*, involves a series of actions for sourcing, making, delivery and return (Li et al. 2011; Long et al. 2012). BI transforms existing data into insights for better planning in all key SCM areas of an organization. In a dynamic market, organizations can assess possible impacts of changing demand - supply conditions and then update existing planned actions based on "what-if analysis" and analytical insights supported by BI (Zhou et al. 2011). BI can also report on historical data such as market growth patterns, customer behaviour changes, and sales trends which help SC managers to *plan* (Sharda et al. 2013). BI supported SCM planning enables managers to become more proactive and efficient in fulfilling uncertain customer needs and thus gain Agility in the SC (Shanks et al. 2010a). SC managers can better respond to changes where those changes can be predicted in advance. BI enabled SCM planning helps managers to be more responsive to market demand by predicting uncertainties and reacting to them before their competitors (Yusuf et al. 2014a). BI enables SC managers to *plan* the appropriate course of action to deal with change.

The second area of SCM is '*source*'. Sourcing involves make or buy decisions, supplier selection, supplier performance evaluation, and inventory management (Luo et al. 2009). BI provides data to managers to analyse supplier key performance indicators and inventory consumption trends during sourcing decision making (Ranjan 2008). In a joint SC network scenario, BI offers integrated data repositories and tools to present such data in a meaningful presentation format to visualize relevant information across the network (Gessner et al. 2009). BI can also be used for ABC analysis and inventory aging analysis to help managers achieve efficiencies in warehouse management. BI supported sourcing helps managers to fulfil customer needs quicker than competitors. In a complex SC network, participating organizations need to know required information about other participating organizations (Power et al. 2001). For example, in order to satisfy a sudden bulk export order, a manager would need to be able to access key information such as ready stocks on hand in finished goods warehouse instantly. If no ready stocks are available, that manager would need to know the stock on hand at the raw material supplier's warehouse and the time required to source those raw materials to determine time to market. Any delay in information availability will impact the SC's quickness (Xun et al. 2009). An integrated real time information dashboard is the key in this scenario. BI can process data from different sources and provide visibility across the SC to respond to the market and act quickly to *source* products to meet customer demand (Sahay et al. 2008).

The third SCM function, '*make*', transforms raw materials into products that meet market demand. (Li et al. 2011). '*Make*' processes deal with capacity management, production scheduling, material flow, manufacturing, packaging and so forth. A central view of demand forecasts, raw material availability and production capacity analysis supported by BI tools helps managers to be well prepared for large scale production (Yusof et al. 2013). BI embedded 'making' supports an organization to become more flexible and to be able to reconfigure resources easily to satisfy customized demands. Some scenarios require flexible production technical flow options to increase productivity. BI can provide data evidence about historical performance of various production configurations (Yusuf et al. 1999). Thus, BI can offer insights for managers to derive and evaluate alternatives when producing goods for the market (Zhang et al. 2013). Therefore, it is expected that the use of BI in this area of SCM improves the ability of SC managers to be flexible and competent in responding to the unexpected changes in the market (Isik et al. 2013).

The fourth SCM area, '*delivery*' hands over finished products to the customer. *Delivery* manages order processing, warehousing, transport management, delivery schedule, delivery time etc. As suggested by Trkman et al. (2010a), there is a practice of outsourcing delivery activities to external logistic companies which forces organizations to have tools which can monitor and analyse the outsourcing providers performance. BI can help SC managers to track picking performance, shipping performance and delivery time records of the third party logistics company to gain operational efficiency and cost effectiveness (Davenport 2006). BI supported decisions help SC managers to attain faster delivery time, delivery cost savings, flexibility and effectiveness (Schläfke et al. 2013). As delivery is a tail ended SC process, any deficiency in delivery process impacts overall SC performance. Thus, decisions supported by BI improves agility not only in the *delivery* process but also in the overall SC.

The final SCM function, '*return*' deals with used or defective products returned by customers (Chen et al. 2011b). Lawton (2008) suggests that only 5% of total returns are due to product defects, the rest being due to not meeting customer tastes or expectations. Thus customer taste and preference data analysis can be a critical input for SCM decision making (Chen et al. 2011b; Xiao et al. 2010). BI can support such complex analysis based on historical consumptions data in different markets to indicate insights on how to reduce returns (Phan et al. 2010). Phillips managed to reduce its *returns* by $100 million per year by utilizing an integrated system that collects and presents data for operational decision making for SCM managers (Tony 2003). *Returns* also influence the wholesale and retail prices as loss due to *return* needs to be considered in the product pricing process (Chen et al. 2011b). BI supported *return* forecast reporting can assist SCM managers in pricing decisions (Bose 2009) and in formulating effective return policy in association with other parties within the SC (Ding et al. 2008). *Return* has significant impact on SC inventory management as there is significant uncertainty about when returns occur and about the quantity of return stocks. Such uncertainty can be reduced by analysing historical sales and return data (de Brito et al. 2009). BI can deliver return analysis reports to relate production defects with customer satisfaction (Xiao et al. 2010). Product performance

analysis, consumer behaviour analysis, and return cost analysis are some examples of BI reporting that assists managers to take corrective actions (Ranjan 2008). BI helps managers to monitor *return* performance to *plan* any possible product recall from market (Xiao et al. 2010).

Based on the above, we conclude with the following propositions:

*P1: BI capability reduces task uncertainty within the key areas of Supply Chain Management (Plan, Source, Make, Deliver, and Return).*

*P2: Reduced task uncertainty improves decision making in the key areas of SCM which in turn results in improved SC Agility*

# 5　Research Methodology

The selection of an appropriate research method together with effective design is the key for data collection to derive meaningful conclusion. The current study is exploratory in nature as we want to explore different uses of BI in SCM key areas especially in a dynamic market context. Existing empirical research related to BI use in business has adopted either case studies (Shanks et al. 2010a) or surveys (Dong et al. 2009; Oliveira et al. 2012; Trkman et al. 2010a) as a research method. As Neuman (2005) suggested, a survey is more related to quantitative research where a number of data samples are collected to explain the relationship, while a case study is an in depth examination of one or multiple units over the periods (Neuman 2005).

This study will adopt a multiple case study approach. Justification for such selection is derived from the argument made by Yin (1994) who suggests that case study research is appropriate when "a 'how' or 'why' question is being asked about a contemporary set of events, over which the investigator has little or no control" (Yin 1994). A case study is selected for this research as it is interested in exploring how BI can be integrated in key SCM processes. This requires in depth study of relevant events in an organization. A multiple case approach has been chosen to explore for patterns in events across multiple case sites.

# 6　Conclusion and Future Study

Achieving an agile SC is challenging because of the complexity involved in SCM. In this paper, we discuss the potential for BI use in SCM processes to assist managers in reducing task uncertainty, and making more informed decisions which ultimately assist in achieving SC agility. Based on our review of the existing literature, we identified a number of gaps in the current understanding regarding how SC Agility is achieved. Some existing studies found that IT can be instrumental in achieving SC Agility. However, IT has been considered in a broad sense in most of those studies. Arguably, different IT components such as ERP, BI, Internet and so on may have different effects on SC Agility. In addition, most of the existing studies cover partial aspects of SCM processes and SC Agility. Hence, a comprehensive study is required focusing on specific IT tools, all key SCM processes and all key components of SC Agility.

A conceptual research model has been proposed in this paper to guide future studies in addressing these identified research gaps. Based on RBV, this study argues that organizations can use BI strategically to create valuable, rare, in-imitable and non-substitutable resources to manage the key areas of SCM that will help achieve SC agility. OIPT has also been used in this study to support our argument that BI can be used in SCM key processes to reduce task uncertainty. Drawing on RBV and OIPT, the proposed research model conceptualises how a BI capability enables organizations to reduce task uncertainties through improved decision making in SCM areas, which in turn leads to improved SC agility. The model also offers an extended scope of study by including five key SCM areas and four possible components of SC agility.

We believe that our proposed research model will help researchers and practitioners to extend the existing knowledge of how SC agility can be achieved. It also highlights the importance of developing a BI capability effectively to reduce task uncertainties in SCM processes and ultimately improve SC agility.

The current study doesn't consider the maturity of BI (Cosic et al. 2015) and organizational influences in the research model. Relative effectiveness of BI use in SCM areas is also not within the scope of this research. These can be considered as limitations of this study and an opportunity for further research.

As next step, case study research involving multiple participants from five organisations will be undertaken. Within each organisation we will interview six to eight managers involved in managing their SC, examine the BI capability within each organization to support the five key areas of SCM and review relevant documentations that are relevant to examine the improvement in each organization's SC agility. Through these in-depth case studies, we will identify mechanisms for effective BI use to create organizational capability to achieve SC Agility.

**COPYRIGHT**